\documentclass[preprint,12pt]{elsarticle} 

\usepackage{graphicx}
\usepackage{amssymb}
\usepackage{lineno}

\usepackage{hyperref}[red]
\usepackage{longtable}
\usepackage{pdflscape}
\usepackage{multirow, booktabs}
\usepackage{multicol}
\usepackage{amsmath}
\usepackage{tikz} 
\usepackage{graphicx}
\usetikzlibrary{positioning}
\usepackage[utf8]{inputenc}
\usepackage{graphicx}
\usepackage{booktabs}
\usepackage{amsthm}
\usepackage{amsmath}
\usepackage{amssymb}
\usepackage{hyperref}[red]
\usepackage{amsfonts}
\usepackage{cuted,float}
\usepackage{makeidx}
\usepackage{amsthm,amscd,mathrsfs,latexsym,amsfonts,amssymb}

\usepackage[left=2cm,right=2cm,top=2cm,bottom=2.5cm]{geometry}
\usepackage{lipsum}
\usepackage{pagecolor}
\usepackage{afterpage}
\theoremstyle{plain}

\journal{Epidemics}

\begin{document}

\begin{frontmatter}


\title{Ebola transmission dynamics: will future Ebola outbreaks become cyclic?}



\author[1,2,3]{David Niyukuri\corref{mycorrespondingauthor}}
\cortext[mycorrespondingauthor]{Corresponding author}
\ead{david.niyukuri@ub.edu.bi}
\author[1,2]{Kelly Jo\"elle Gatore Sinigirira}
\author[4]{Jean De Dieu Kwizera}
\author[5]{Salma Omar Abd-Almageid Adam}

\address[1]{Department of Mathematics, University of Burundi, Burundi}
\address[2]{Doctoral School, University of Burundi, Burundi}
\address[3]{The South African Department of Science and Technology--National Research Foundation (DST-NRF) Centre of Excellence in Epidemiological Modelling and Analysis (SACEMA), Stellenbosch University, Cape Town, South Africa}
\address[4]{Institut Superieur des Cadres Militaires, Bujumbura, Burundi}
\address[5]{Department of Mathematics, College of Computer Sciences and Mathematics, University of Bahri, Karthoum, Sudan}

\begin{abstract}

The Ebola Virus Disease (EVD) can persist in some body fluids after clinical recovery. In Guinea and the Democratic Republic of the Congo, there are well-documented cases of EVD re-emergence that were associated with previous outbreaks. In many cases, male EVD survivors were associated with the re-introduction of new outbreaks. This has shown that even after controlling an EVD outbreak, a close biomedical monitoring of survivors and contacts is critical to avoiding future outbreaks. Thus, in order to explore the main features of EVD transmission dynamics in the context of re-emergence, we used a compartmental model by considering vaccination around EVD contacts. Analytical and numerical analyses of the model were conducted. The model is mathematically and epidemiologically well-posed. We computed the reproductive number ($R_0$) and the disease equilibrium points (disease-free equilibrium and endemic equilibrium) for the re-emerging outbreak. The stability analysis of the model around those equilibrium points was performed. The model undergoes a backward bifurcation at $R_0$ close to 1, regardless $R_0<1$, the disease will not be eradicated. This means that R0 cannot be considered as an intervention control measure in our model. 


\end{abstract}


\end{frontmatter}


\section{Introduction}

It is known that the Ebola Virus Disease (EVD) infection can persist in some body fluids after clinical recovery and clearance of virus from the bloodstream. In a systematic review, authors highlighted that EVD viral RNA can still be detected in semen (day 272), aqueous humor (day 63), sweat (day 40), urine (day 30), vaginal secretions (day 33), conjunctival fluid (day 22), faeces (day 19) and breast milk (day 17) \cite{chughtai2016persistence}.Thus, this has important implication on EVD transmission dynamics and control strategies. In the past, sexual transmission from recovered Ebola survivors has been documented \cite{thorson2016systematic, abbate2016potential}. \par

During the EVD outbreak in West Africa in 2014-2015, the World Health Organization (WHO) declared the end of Ebola virus (EBOV) transmission in the Republic of Guinea on 29 December 2015, a resurgence of EVD cases was reported in February 2016 almost after 2 months \cite{diallo2016resurgence, subissi2018ebola}. Surprisingly, analysis showed that this resurgence in Guinea was linked to previous outbreak of 500 days ago. In a longitudinal study in Liberia of a cohort of EVD survivors and their close contacts to understand EVD sequelae showed that the EVD viral RNA in semen persisted for a maximum of 40 months \cite{prevail2019longitudinal}. Recent EVD outbreaks in the Democratic Republic of Congo (DRC) and Guinea in 2021, were close to previous outbreaks in those respective countries.\\

In the Democratic Republic of the Congo (DRC), on 7 February 2021, an Ebola virus disease (EVD) outbreak was declared by the Ministry of Health (MOH) of the DRC. On 8 October 2021, the Ministry of Health announced an outbreak of Ebola virus disease (EVD) in Beni Health Zone, North Kivu Province. According to phylogenetic analysis, these sporadic outbreaks were linked to the 2018-2020 Nord Kivu/Ituri EVD outbreak, hence they were not new spillover events \cite{oct2021virological}.

In the Republic of Guinea, on 14 February 2021, Guinean health authorities declared an EVD outbreak, the lineage shows considerably lower divergence than would be expected during sustained human-to-human transmission, which suggests a persistent infection with reduced replication or a period of latency \cite{keita2021resurgence}. A phylogenetic tree shows the new virus falls between virus samples from the 2013–16 epidemic \cite{tiper2022tracking}. Thus, the virus which caused recent outbreaks barely differs from the strain seen 5 to 6 years ago, and this becomes a puzzle which led to different hypothesis. The biology of Ebola virus does not support if the virus could lay dormant for 5 years in a survivor of the epidemic all that time. Dan Bausch from the United Kingdom’s Public Health Rapid Support Team argued that \textit{“From the tree, you’d conclude that it is a virus that persisted in some way in the area, and sure, most likely in a survivor”}. He added that it is also hard to rule out scenarios such as a small, unrecognized chain of human-to-human transmission. “For example, a 2014 survivor infects his wife a few years after recovery, who infects another male, who survives and carries virus for a few years, then infecting another women, who is then seen by a nurse who dies”—the index case in the new outbreak, he said \cite{kupferschmidt2021new}.

\begin{figure}[!h]
\centering
\includegraphics[width=0.6\linewidth]{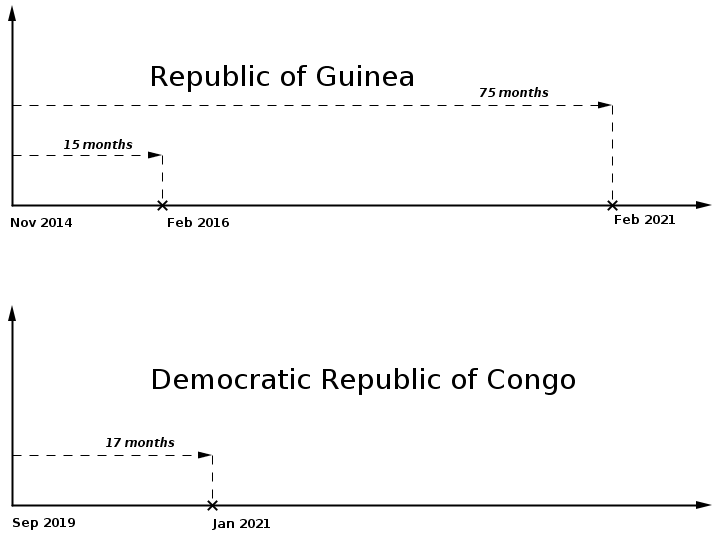}
\caption{Re-introduction of EVD in Guinea and the Democratic Republic of Congo. In Guinea, the Current EVD outbreak was identified to be close to Ebola sequence samples in infected individuals in 2014. Same scenario occurred in February 2016, a re-introduction of EVD from a man survivor was reported in Guinea. In the Democratic Republic of Congo, the recent EVD outbreak was associated to the first case which was a woman married to an EVD survivor of 2019.}
\label{fig:resurgence}
\end{figure}

From those reported cases, we can see that EVD epidemics can exhibit transmission cycles in which survivors play a pivotal role for outbreak re-emergence. From above mentioned reported cases, sexual transmission was the main transmission mode for new EVD outbreak re-introduction. Thus, EVD is becoming an emerging sexually transmissible infection of great concerns \cite{surani2018ebola}. Thus, the overall aim of this modelling study is to get insights on which conditions EVD outbreaks will re-emerge from previous outbreaks. Given the effectiveness of the vaccine for Zaire strains which caused the above mentioned outbreaks, we will also determine the value of the reproductive number (R0), the disease-free equilibrium (DFE), and endemic equilibrium (EE) \cite{anggriani2015existence, rafiq2020reliable}, for the re-emerging outbreak. 



\section{Model description and formulation}

\begin{figure}[!ht]
\centering
\includegraphics[width=0.80\linewidth]{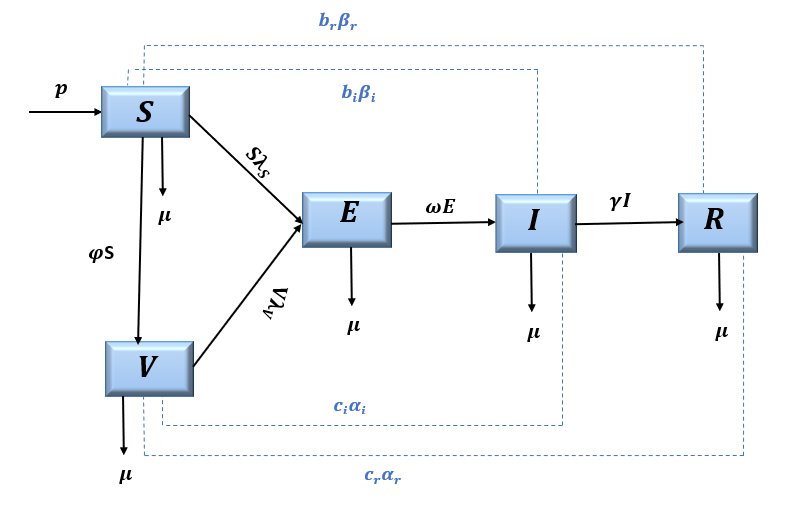}
\caption{Schematic representation of mathematical model of EVD transmission.}
\label{fig:sveir}
\end{figure}

To study the transmission cycles of EVD dynamics, we suggested a Susceptible-Vaccinated-Exposed-Infected-Recovery (SVEIR) type model with 5 compartments which are described in Figure \eqref{fig:sveir}, consisting of coupled nonlinear differential equations \cite{rafiq2020reliable, boujakjian2016modeling}.In our model formulation, we assume that the very first infection of Ebola virus might be a spillover from wild animals to human subjects. We also assume that recovered individuals can trig new infections as it has been reported in DRC and Guinea.   We perform analytical analysis to look at the existence of solutions, and stability analysis of the equilibria of the model. 

Explicitly, we have the class of Susceptible (S) humans at time instant t, this is a class of individuals who are not yet infected but can get infected by the Ebola virus; Given the availability of EVD vaccination, it is feasible to consider a specific vaccinated class (V) which is a class of individuals who are vaccinated. Exposed (E) individuals, they are individuals who were susceptible and vaccinated and  have become exposed after being in contact with any of Ebola infected person of contaminated body fluids or fomites (infected inanimate objects and surfaces) \cite{chughtai2016persistence, bausch2007assessment}, these people  are placed in this class which consists of exposed humans at time instant t; Infected (I) humans at time instant t, exposed who develop the disease; and Recovered (R) class is for infected individuals who survive and finally recover or acquire immunity due to vaccination.


The mathematical model of EVD transmission dynamic is given by the instantaneous change of the flow between compartment, which is described by the following system of non-linear differential equations \ref{model_System1} which assumes $\lambda_s=\frac{1}{N}(b_i\beta_iI+b_r\beta_rR)$ and $\lambda_v=\frac{1}{N}(c_i\alpha_iI+c_r\alpha_rR)$.

\begin{eqnarray}
\left \{
\begin{array}{rcl}
\frac{dS}{dt}&=&p-\frac{S}{N}(b_i\beta_i I+b_r\beta_r R)-(\mu+\varphi)S\\\\

\frac{dV}{dt}&=&\varphi S-\frac{V}{N}(c_i\alpha_i I+c_r\alpha_r R)-\mu V\\\\

\frac{dE}{dt}&=&\frac{S}{N}(b_i\beta_i I+b_r\beta_r R)+\frac{V}{N}(c_i\alpha_i I+c_r\alpha_r R)-(\mu+\omega)E\\\\

\frac{dI}{dt}&=&\omega E-(\gamma+\mu+\delta)I\\\\
\frac{dR}{dt}&=&\gamma I-\mu R
\end{array}
\right. 
\label{model_System1}
\end{eqnarray}

with $N(t)=S(t)+V(t)+E(t)+I(t)+R(t)$, which means that the population change over time.



\begin{table}[!ht]
\caption{Parameters for the S-V-E-I-R Ebola Virus Disease dynamic model}
\begin{center}
\begin{tabular}{l|p{9cm}}
\hline
  Parameter & Definition \\
   \hline
   $p$ & Recruitment rate of individuals into the population  \\
   $\varphi$ & Proportions of susceptible individuals who are vaccinated\\
   $\mu$ & Natural death rate\\
   $b_i$ & contact rate between infected individuals and susceptible individuals\\
    $\beta_i$ &  probability for susceptible individuals of being in contact with infected individuals that\\
    & occurs an infection.\\
   $b_r$ &  contact rate between recovered individuals and susceptible individuals\\
   $\beta_r$ & probability for susceptible individuals of being in contact with recovered individuals that \\
    & occurs an infection.\\
   $c_i$ & contact rate between infected individuals and vaccinated individuals\\
   $\alpha_i$ & probability for vaccinated individuals of being in contact with infected individuals that\\
   & occurs an infection.\\
   $c_r$ & contact rate between recovered individuals and vaccinated individuals\\
   $\alpha_r$ & probability for vaccinated individuals of being in contact with recovered individuals that \\
   & occurs an infection.\\
   $\omega$ & Exit rate from exposed class.\\
   $\gamma$ & Exit rate from infected class.\\
   $\delta$ & disease induced death rate\\
   \hline
\end{tabular}
\end{center}
\label{table_params}
\end{table}

New EVD infections are results of the interaction between susceptible or vaccinated and infected individuals respectively at rate $\beta_i$ and $\alpha_i$, this gives the rate at which individuals are being exposed to the infection. We also assume that individuals who are considered recovered can interact with susceptible and vaccinated individuals respectively at rate $\beta_r$ and $\alpha_r$ hence susceptible individuals become exposed by interacting with EVD survivors, and this can cause re-introduction of EVD. This is supported by current evidence of resurgence or re-introduction of EVD through sexual intercourse. The rate of increase or decrease of susceptible population is given by $p$, which can be by natural deaths, migrations or new births. Individuals die at rate $\mu$ by natural deaths. EVD death rate is given by $\delta$. The incubation period also known as latent period which is the time between infection and sickness is given by $\omega$. All parameters of the SVEIR model are given in Table \ref{table_params}.

We have seen that new introduction of EVD infections have been driven by men EVD survivors. Thus, the re-introduction parameter $\alpha_r$ and $\beta_r$ should depend also on proportion of sexually active men who survived the infection. To solve the system of nonlinear differential equations, we used the numerical method Runge–Kutta of order 4 (RK4). To fit the model, and estimate some parameters of the model, we used record data of reported EVD cases in the Republic of Guinea and DRC.

\section{Model analysis}

The analytical analysis of the systems of ODEs \eqref{model_System1} shows the dynamical behaviour of the solutions of the system. With the intrinsic nature of transmission dynamic through contact, we will estimate the basic reproduction number R0, analyze different equilibria of the model \cite{anggriani2015existence,rafiq2020reliable,madubueze2018global,tahir2018ebola}, at the disease-free equilibrium (DFE), and endemic equilibrium (EE) points. We will also analyze the attainment  of herd immunity.

\subsection{Bounded and invariant region}
The positivity and boundness of the solution are investigated to show that the model is biologically and mathematically well posed.\\

\textbf{Theorem:} The feasible region $\Omega$ defined by:

\begin{eqnarray}
\Omega &=&\{S(t);V(t),E(t),I(t),R(t)\in\mathbb{R}_+^5:N(t)\leq\dfrac{p}{\mu}\}
\end{eqnarray}

with $S(0),V(0),E(0),I(0),R(0)\leq 0$ is positively invariant.

\begin{proof}
Adding the equation of the system \ref{model_System1}, We get:
\begin{eqnarray}
\dfrac{dN}{dt}&\leq& p-\mu N
\label{equation0}
\end{eqnarray}
Solving \ref{equation0}, We have:
\begin{eqnarray*}
0\leq N(t)\leq\dfrac{p}{\mu}+N(0)e^{-\mu t}
\end{eqnarray*}
$N(0)$ is the initial values of the total population $N(t)$  and
\begin{eqnarray}
\lim\limits_{t \rightarrow +\infty} sup N(t)\leq \dfrac{p}{\mu}
\end{eqnarray}
So $\Omega=\{S(t);V(t),E(t),I(t),R(t)\}$ is positively invariant, and the right handside of equations of the model \eqref{model_System1} is continuous with continuous partial derivative in $\Omega$. Hence, there is a unique solution.
\end{proof}

\subsection{Disease-free equilibrium}

The disease-free equilibrium (DFE) is the state at which there is no infectious in the population.\\
So,
$E=I=R=0$ and $ \frac{dS}{dt}=\frac{dV}{dt}=\frac{dE}{dt}=\frac{dI}{dt}=\frac{dR}{dt}=0 $
Then, the model admit the disease free equilibrium point $E_{dfe}=(S_0,V_0,0,0,0)$ with
\begin{eqnarray}
S_0=\frac{p}{\mu+\varphi}\mbox{  and  } V_0=\frac{p\varphi}{\mu(\mu+\varphi)}
\end{eqnarray}

\subsection{Basic reproduction number}

We now determine an important threshold parameter in epidemic modelling known as the basic reproduction number $(R_0)$, it gives the average number of generated infection cases by an infectious individual. Using the New Generation Number described in (references), we will compute the basic reproduction number $R_0$.
Let's define the transmission matrix $f$ and $v$
\begin{eqnarray}
f=
\begin{pmatrix}
\lambda_s S+\lambda_v \\
0\\
0 
\end{pmatrix}
\mbox{  and  }
v=
\begin{pmatrix}
(\mu + \omega)E 
\\-\omega E + (\gamma+\mu+\delta)I\\
-\gamma I+\mu R 
\end{pmatrix}
\end{eqnarray}

The corresponding Jacobian matrix $F$ and $V$ of $f$ and $v$ evaluated at the DFE are respectively:


\begin{eqnarray}
F_{E_{dfe}}=
\begin{pmatrix}
0&\dfrac{\mu}{\mu+\varphi} (b_i\beta_i+\dfrac{c_i\alpha_i\varphi}{\mu})&\dfrac{\mu}{\mu+\varphi}(b_r\beta_r + \dfrac{c_r\alpha_r\varphi}{\mu})\\0&0&0\\0&0&0 
\end{pmatrix}
\mbox{  and  } 
V=
\begin{pmatrix}
\mu + \omega &  & 0\\
-\omega & (\gamma+\mu+\delta) &0\\
0&-\gamma &\mu
\end{pmatrix}
\end{eqnarray}
 $FV^-{1}$ is the next generation number given by 

 \newpage

\begin{eqnarray*}
\left(
\begin{array}{c}
\dfrac{\omega(\mu b_i\beta_i+\varphi c_i\alpha_i)}{(\mu+\varphi)(\mu+\omega)(\mu+\gamma+\delta)}+\dfrac{\omega\gamma(\mu b_r\beta_r+\varphi c_r\alpha_r)}{\mu (\mu+\varphi)(\mu+\omega)(\mu+\gamma+\delta)}\\
 0 \\
 0 
\end{array}
\right.
\end{eqnarray*}
\begin{eqnarray}
\left.
\begin{array}{ccc}

  \dfrac{(\mu b_i\beta_i+\phi\alpha_i c_i)}{(\mu+\varphi)(\mu+\gamma+\delta)}+\dfrac{(\mu b_r\beta_r+\varphi c_r \alpha_r)}{\mu(\mu+\varphi)(\mu+\omega)(\mu+\gamma+\delta)}  &\dfrac{1}{\mu(\mu+\varphi)}(\mu b_r\beta_r+\varphi c_r\alpha_r)  \\
    0  &  0\\
    0  &  0
\end{array}
\right)
\end{eqnarray}

It follows that the $R_0$ is given by
$R_0 = \rho(FV^{-1})$ Where $\rho$ is the spectral radius of the new generation infection\\
So, the basic reproduction number becomes
\begin{eqnarray}\label{r_not}
R_0=\rho FV^{-1}=\dfrac{\omega[\mu(\mu b_i\beta_i+ c_i\alpha_i\varphi)+\gamma(\mu b_r\beta_r+c_r\alpha_r\varphi)]}{\mu(\mu+\varphi)(\mu+\omega)(\mu+\gamma+\delta)}
\end{eqnarray}

\subsection{Local stability of DFE} 
\textbf{Theorem}: The disease free equilibrium point, $E_{dfe}$ is locally asymptotically stable if $R_0 < 1$ and unstable if $R_0 > 1$.
\begin{proof}
Using the Jacobian matrix of Equations[Ref], we established the local stability of the disease-free equilibrium and evaluated it at disease free equilibrium $E_{dfe}$.This is accomplished by evaluating Equations [Ref]'s Jacobian matrix at $E_{dfe}=(\dfrac{p}{\mu+\varphi};\dfrac{p\varphi}{\mu(\mu+\varphi)};0;0,0)$ 
The eigenvalues of the Jacobian matrix of the model equations at $E_{dfe}$ are used to determine the model's local stability. As a result, the Jacobian matrix of Equations [Ref] is given by

\begin{eqnarray}
J_{E_{dfe}}=
\begin{pmatrix}
-(\mu + \varphi) & 0& 0& -\dfrac{\mu b_i\beta_i}{\mu+\varphi}&-\dfrac{\mu b_r\beta_r}{\mu+\varphi}\\\\
\varphi&-\mu&0&-\dfrac{\varphi c_i\alpha_i}{\mu+\varphi}& -\dfrac{\varphi c_r\alpha_r}{\mu+\varphi}\\\\
0&0&-(\mu+\omega)&\dfrac{\mu b_i\beta_i+\varphi c_i\alpha_i}{\mu+\varphi}&\dfrac{\mu b_r\beta_r+\varphi c_r\alpha_r}{\mu+\varphi}\\\\
0&0&\omega & -(\gamma+\mu+\delta) &0\\\\
0&0&0&-\gamma &-\mu
\end{pmatrix}
\label{matrice1}
\end{eqnarray}

The jacobian matrix have one column with diagonal entries.$-\mu$ are eigenvalue of the matrix~\eqref{matrice1}. Eliminating the second columns and the corresponding row, we have:\\

\begin{eqnarray}
\begin{pmatrix}
-(\mu + \varphi) & 0 & -\dfrac{\mu b_i\beta_i}{\mu+\varphi}&-\dfrac{\mu b_r\beta_r}{\mu+\varphi}\\
0&-(\mu+\omega)&\dfrac{\mu b_i\beta_i+\varphi c_i\alpha_i}{\mu+\varphi}&\dfrac{\mu b_r\beta_r+\varphi c_r\alpha_r}{\mu+\varphi}\\
0&\omega & -(\gamma+\mu+\delta) &0\\
0&0&-\gamma &-\mu
\end{pmatrix}
\label{matrice2}
\end{eqnarray}
Eliminating the first column, we have $-(\mu+\varphi$)as eigenvalue of the matrix~\eqref{matrice2}. So the Jacobian matrix become:
\begin{eqnarray}
\begin{pmatrix}
-(\mu+\omega)&\dfrac{\mu b_i\beta_i+\varphi c_i\alpha_i}{\mu+\varphi}&\dfrac{\mu b_r\beta_r+\varphi c_r\alpha_r}{\mu+\varphi}\\
\omega & -(\gamma+\mu+\delta) &0\\
0&-\gamma &-\mu
\end{pmatrix}
\label{matrice3}
\end{eqnarray}

Assume that:
\begin{eqnarray*}
J_1&=&\mu+\omega\\
J_2&=&(\gamma+\mu+\delta)\\
J_3&=&\mu\\
A &=& \mu b_i\beta_i+\varphi c_i\alpha_i\\
B &=& \mu b_r\beta_r+\varphi c_r\alpha_r\\
C&=&\mu+\varphi
\end{eqnarray*}
The remaining eigenvalues can be calculated  by computing all the solutions of the characteristic equation $|J_{dfe}-\lambda I|=0$\\
So, 
\begin{eqnarray}
|J_{dfe}-\lambda I|=\left|
\begin{matrix}
 -(J_1+\lambda) & \dfrac{A}{C} &\dfrac{B}{C} \\
 \omega & -(J_2+\lambda)   & 0\\
0&\gamma & -(J_3+\lambda)
\end{matrix}
\right|
\end{eqnarray}
\label{matrice4}
\begin{eqnarray}
(J_1+\lambda)(J_2+\lambda)(J_3+\lambda)-\omega\gamma \dfrac{B}{C}-(J_3+\lambda)\omega\dfrac{A}{C} =0
\label{equation1}
\end{eqnarray}
Thus, the equation~\eqref{equation1} can be written as a polynomial of degree 3:
\begin{eqnarray}
\lambda^3+A_2\lambda^2+A_1\lambda+A_0=0 
\label{equation2}
\end{eqnarray}
Where
\begin{eqnarray}
A_2 &=&J_1+J_2+J_3 >0\\ \nonumber
A_1 &=&J_1J_2+J_1J_3+J_2J_3-\omega\dfrac{A}{B}>0\mbox{ with }(\omega\dfrac{A}{B}< J_1J_2+J_1J_3+J_2J_3)\\   \nonumber
A_0&=& J_1J_2J_3-\omega(J_3\dfrac{A}{C}+\gamma\dfrac{B}{C})\\\nonumber
    &=&J_1J_2J_3C(1-R_0) >0 \mbox{ if } R_0<1\nonumber
\end{eqnarray}

According to Descarte'rules of sign \cite{anderson1998descartes},it has been observed that all the coefficient of \eqref{equation2} are strictly positive if $R_0<1$. Then, all it's eigenvalue have a negative really part.\\
$E{dfe}$ is locally asymptotically stable by the Poincare Lyapnov theorem \cite{tsokos1969classical}
\end{proof}

\subsubsection{Global stability of DFE}
\textbf{Theorem}: The disease free equilibrium point, $E_{dfe}$ is globally asymptotically stable if $R_0 \leq 1$ and unstable if $R_0 > 1$.\\
\begin{proof} Let's define
\begin{equation}
    L=(E,I,R)=aE+bI+cR
\end{equation}
as a Lyapunov function for some non negative values of $a$, $b$, and $c$. The time derivative of $L$ is given by
\begin{eqnarray*}
\dfrac{dL}{dt}&=&a\dfrac{dE}{dt}+b\dfrac{dI}{dt}+c\dfrac{dR}{dt}\\
        &\leq&a\Big[\dfrac{\mu}{\mu+\varphi}(b_i\beta_iI+b_r\beta_rR)+\dfrac{\varphi}{\mu+\varphi}(c_i\alpha_iI+c_r\alpha_rR)-(\mu+\omega)E\Big]\\
        & & + b[\omega E-(\gamma+\mu+\delta)I]+c(\gamma I+\mu R)\\
        &=&[-a(\mu+\omega)+b\omega]E+\big[a(\dfrac{\mu}{\mu+\varphi}b_i\beta_i+\dfrac{\varphi}{\mu+\varphi}c_i\alpha_i)-b(\gamma+\mu+\delta)+c\gamma\big]I+\\
        &&\big[a(\dfrac{\mu}{\mu+\varphi}b_r\beta_r+
        \dfrac{\varphi}{\mu+\varphi}c_r\alpha_rI)-c\mu \big]R
\end{eqnarray*}
Computing the Lyapunov function coefficients so that the coefficients of $E$, $I$, $R$, are all equal to zero yields to
\begin{eqnarray}
\left\{
\begin{array}{l}
a=\dfrac{\omega}{\mu+\omega}\\\\
b=\dfrac{\omega[\mu(\mu b_i\beta_i+\varphi c_i \alpha_i)+\gamma(\mu b_r\beta_r+\varphi c_r\alpha_r)]}{\mu(\mu+\varphi)(\mu+\omega)(\gamma+\mu+\delta)}\\\\
c=\dfrac{\omega(\mu b_r\beta_r+\varphi c_r \alpha_r)}{\mu(\mu+\varphi)(\mu+\omega)}
\end{array}
\right.
\label{systeme5}
\end{eqnarray}
The time derivative of the Lyapunov function may be written using these coefficients as

\begin{eqnarray*}
\dfrac{dL}{dt}&=&\dfrac{\omega}{\mu+\omega}\dfrac{dE}{dt}+\dfrac{\omega[\mu(\mu b_i\beta_i+\varphi c_i \alpha_i)+\gamma(\mu b_r\beta_r+\varphi c_r\alpha_r)]}{\mu(\mu+\varphi)(\mu+\omega)(\gamma+\mu+\delta)}\dfrac{dI}{dt}+\dfrac{\omega(\mu b_r\beta_r+\varphi c_r \alpha_r)}{\mu(\mu+\varphi)(\mu+\omega)}\dfrac{dR}{dt} \\
&\leq&\dfrac{\omega}{\mu+\omega}\Big[\dfrac{\mu}{\mu+\varphi}(b_i\beta_iI+b_r\beta_rR)+\dfrac{\varphi}{\mu+\varphi}(c_i\alpha_iI+c_r\alpha_rR)-(\mu+\omega)E\Big]\\
&&+R_0[\omega E-(\gamma+\mu+\delta)]+\dfrac{\omega(\mu b_r\beta_r+\varphi c_r \alpha_r)}{\mu(\mu+\varphi)(\mu+\omega)}(\gamma I+\mu R)\\
&=&\omega(R_0-1)E
\end{eqnarray*}
when $R_0<1$ or $R_0=1$, it is evident that $\dfrac{dL}{dt}\leq 0$. Furthermore $\dfrac{dL}{dt}=0$ iff $E=I=R=0$ and the largest compact invariant set in $\omega$ such taht $\dfrac{dL}{dt}\leq 0$ is the $E_{dfe}$.\\
As a result, according to the Lyapunov-Lasalle Invariance Principle \cite{huang2010global}, the disease free equilibrium is globally stable in $\Omega$  if  $R_0\leq 0$

\end{proof}
\newpage

\subsection{Existence of endemic equilibrium}

The endemic equilibrium (EE) is defined as the point at which the disease remains present in the population. To compute it, we consider $\dfrac{dS}{dt}=\dfrac{dV}{dt}=\dfrac{dE}{dt}=\dfrac{dI}{dt}=\dfrac{dR}{dt}=0$ and
\begin{eqnarray}
\left \{
\begin{array}{rcl}
p-\lambda_s S^*-(\mu+\varphi)S^*&=&0\\\\
\varphi S^*-\lambda_v V^* -\mu V^*&=&0\\\\
\lambda_s S^*+\lambda_v V^*-(\mu+\omega)E^*&=&0\\\\
\omega E^*-(\gamma+\mu+\delta)I^*&=&0\\\\
\gamma I^*-\mu R^*&=&0
\end{array}
\right. 
\label{model_System}
\end{eqnarray}

We obtain 
\begin{eqnarray}
\left \{
\begin{array}{rcl}
S^*&=&\dfrac{p}{\lambda_s+(\mu+\varphi)}\\\\
V^*&=&\dfrac{p\varphi}{[\lambda_s+(\mu+\varphi)](\lambda_v+\mu)}\\\\
E^*&=&\dfrac{\lambda_sp(\lambda_v+\mu)+\lambda_v\varphi p}{(\mu+\omega)[\lambda_s+(\mu+\varphi)](\lambda_v+\mu)}\\\\
I^*&=&\dfrac{\omega[\lambda_s p(\lambda_v+\mu)+\lambda_v\varphi]}{(\mu+\omega)(\gamma+\mu+\delta)[\lambda_s+(\mu+\varphi)](\lambda_v+\mu)}\\\\
R^*&=&\dfrac{\gamma\omega[\lambda_s p(\lambda_v+\mu)+\lambda_v\varphi p]}{\mu(\mu+\omega)(\gamma+\mu+\delta)[\lambda_s+(\mu+\varphi)](\lambda_v+\mu)}\\\\
\end{array}
\right. 
\label{model_System}
\end{eqnarray}

\textbf{\subsection{Local stability of EE}}
We can calculate the stability of the endemic point using the eigenvalue approach and evaluate it at $E^*$, but our system will require sophisticated computations, which could lead to many errors. Then, at the DFE ~\cite{berezovskaya2004simple}, we apply the center Manifold theory created by Chavez and Song for bifurcation analysis.\\
From system~\eqref{model_System1}, let us rewritten the variables  $S,V,E,I,R,$ in the dimensionless states as $S=x_1,V=x_2,E=x_3,I=x_4,R=x_5$ and let ${X=(x_1,x_2,x_3,x_5 x_5)^T}$,
the expressions of $X$ in vector notation.  \\
The model system ~\eqref{model_System1} can be rewritten in terms of  
$\dfrac{dX}{dt}=(f_1, f_2, f_3,\cdots f_5)^T$ \\
where
\begin{eqnarray}
\left \{
\begin{array}{rcl}
\dfrac{dx_1}{dt}&=&p-\dfrac{x_1}{N}(b_i\beta_i x_4+b_r\beta_r x_5)-(\mu+\varphi)x_1\\\\
\dfrac{dx_2}{dt}&=&\varphi x_1-\dfrac{x_2}{N}(c_i\alpha_i x_4+c_r\alpha_r x_5)-\mu x_2\\\\
\dfrac{dx_3}{dt}&=&\dfrac{x_1}{N}(b_i\beta_i x_4+b_r\beta_r x_5)+\dfrac{x_2}{N}(c_i\alpha_i x_4+c_r\alpha_r x_5)-(\mu+\omega)x_3\\\\
\dfrac{dx_4}{dt}&=&\omega x_3-(\gamma+\mu+\delta)x_4\\\\
\dfrac{dx_5}{dt}&=&\gamma x_4-\mu x_5
\end{array}
\right. 
\label{Systemwithx}
\end{eqnarray}
and Where $N=x_1+x_2+x_3+x_4+x_5$\\

This can be used to show that when $R_0=1$, the bifurcation parameter is $\beta_i=\beta_i^*$ , it follows that
\begin{eqnarray}
1=\dfrac{\omega[\mu(\mu b_i\beta_i+ c_i\alpha_i\varphi)+\gamma(\mu b_r\beta_r+c_r\alpha_r\varphi)]}{\mu(\mu+\varphi)(\mu+\omega)(\mu+\gamma+\delta)}
\end{eqnarray}
so
\begin{eqnarray}
\beta_i^*=\dfrac{\mu(\mu+\varphi)(\mu+\omega)(\mu+\gamma+\delta)-\omega\Big[\gamma(\mu b_r\beta_r+c_r\alpha_r\varphi)+ c_i\alpha_i\varphi\Big]}{\omega \mu^2 b_i}
\end{eqnarray}
The center manifold theory will aid in the description of the system's dynamics at $\beta_i^*$, and the jacobian matrix of system \eqref{Systemwithx} at the DFE $E_{dfe}$ is provided by:
\begin{eqnarray}
J=
\begin{pmatrix}
-(\mu + \varphi) & 0& 0& -\dfrac{\mu b_i\beta_i}{\mu+\varphi}& -\dfrac{\mu b_r\beta_r}{\mu+\varphi}\\\\
\varphi&-\mu&0&-\dfrac{\varphi c_i\alpha_i}{\mu+\varphi}& -\dfrac{\varphi c_r\alpha_r}{\mu+\varphi}\\\\
0&0&-(\mu+\omega)&\dfrac{\mu b_i\beta_i+\varphi c_i\alpha_i}{\mu+\varphi}&\dfrac{\mu b_r\beta_r+\varphi c_r\alpha_r}{\mu+\varphi}\\\\
0&0&\omega & -(\gamma+\mu+\delta) &0\\\\
0&0&0&\gamma &-\mu
\end{pmatrix}
\label{matrice5}
\end{eqnarray}
Assume that: \\
$J_1=\mu+\varphi$, $J_2=\mu$,$J_3=\mu+\omega$,$J_4=\gamma+\mu+\delta$,$J_5=\mu$\\ 
 The right eigenvector corresponding to the zero eigenvalue $W=(w_1,w_2,w_3,w_4,w_5)^T$, is computed using $J.W=0$ which gives  :
\begin{eqnarray}
J=
\begin{pmatrix}
-(\mu + \varphi)&0&0& -\dfrac{\mu b_i\beta_i}{\mu+\varphi}& -\dfrac{\mu b_r\beta_r}{\mu+\varphi}\\\\
\varphi&-\mu&0&-\dfrac{\varphi c_i\alpha_i}{\mu+\varphi}& -\dfrac{\varphi c_r\alpha_r}{\mu+\varphi}\\\\
0&0&-(\mu+\omega)&\dfrac{\mu b_i\beta_i+\varphi c_i\alpha_i}{\mu+\varphi}&\dfrac{\mu b_r\beta_r+\varphi c_r\alpha_r}{\mu+\varphi}\\\\
0&0&\omega & -(\gamma+\mu+\delta) &0\\\\
0&0&0&\gamma &-\mu
\end{pmatrix}
\begin{pmatrix}
 w_1\\\\
 w_2\\\\
 w_3\\\\
 w_4\\\\
 w_5\\\\
 \label{matrice6}
\end{pmatrix}
\end{eqnarray}
we deduce
 
\begin{eqnarray}
\left\{
\begin{array}{l}
-J_1 w_1-\dfrac{\mu b_i\beta_i}{\mu+\varphi}w_4-\dfrac{\mu b_r\beta_r}{\mu+\varphi}w_5=0\\
\varphi w_1-J_2w_2+\dfrac{\varphi c_i\alpha}{\mu+\varphi}w_4-\dfrac{\varphi c_i\alpha}{\mu+\varphi}w_5=0\\
-J_3w_3+\dfrac{\mu b_i\beta_i+\varphi c_i\alpha}{\mu+\varphi}w_4+\dfrac{\mu b_r\beta_r+\varphi c_i\alpha}{\mu+\varphi}w_5=0\\
\omega w_3-J_4w_4=0\\
\gamma w_4-J_5 w_5=0
\end{array}
\right.
\label{systeme_3}
\end{eqnarray}
By setting $W_5 =1$ and solving \eqref{systeme_3} we get:

\begin{eqnarray}
\left\{
\begin{array}{rcl}
w_1&=&-\dfrac{(\gamma b_i\beta_i+\mu b_r\beta_r)]}{(\mu+\varphi)^2}\\
w_2&=&-\dfrac{\varphi\Big[\mu c_i\alpha_i+\gamma c_r\alpha_r-\gamma(\mu+\varphi)^]}{\gamma\mu(\mu+\varphi)}\\
w_3&=&\dfrac{\mu(\gamma+\mu+\delta)}{\gamma\omega}\\
w_4&=&\dfrac{\mu}{\gamma}\\
w_5&=&1
\end{array}
\right.
\label{systeme_4}
\end{eqnarray}
$J$ has also a left eigenvector $V=(v_1,v_2,v_3,v_4,v_5)$ associated with the zero eigenvalue and satisfying $V.J=0$. To find this, we use the jacobian matrix \eqref{matrice5}.\\
So,   
\begin{eqnarray}
J=
\begin{pmatrix}
 v_1&v_2&w_3&v_4&v_5
\end{pmatrix}
\begin{pmatrix}
-(\mu + \varphi)&0&0& -\dfrac{\mu b_i\beta_i}{\mu+\varphi}& -\dfrac{\mu b_r\beta_r}{\mu+\varphi}\\\\
\varphi&-\mu&0&-\dfrac{\varphi c_i\alpha_i}{\mu+\varphi}& -\dfrac{\varphi c_r\alpha_r}{\mu+\varphi}\\\\
0&0&-(\mu+\omega)&\dfrac{\mu b_i\beta_i+\varphi c_i\alpha_i}{\mu+\varphi}&\dfrac{\mu b_r\beta_r+\varphi c_r\alpha_r}{\mu+\varphi}\\\\
0&0&\omega & -(\gamma+\mu+\delta) &0\\\\
0&0&0&\gamma &-\mu
\end{pmatrix}
\label{matrice7}
\end{eqnarray}
Which yields to
\begin{eqnarray}
\left\{
\begin{array}{l}
-J_1 v_1+\varphi v_2=0\\
-J_2v_2=0\\
-J_3v_3+\omega v_4=0\\
-\dfrac{\mu b_i\beta_i}{\mu+\varphi}v_1-\dfrac{\varphi c_i\alpha_i}{\mu+\varphi}v_2+\dfrac{\mu b_i\beta_i+\varphi c_i\alpha_i}{\mu+\varphi}v_3-J_4v_4-\gamma v_5=0\\
-\dfrac{\mu b_r\beta_r}{\mu+\varphi}v_1-\dfrac{\varphi c_r\alpha_r}{\mu+\varphi}+\dfrac{\mu b_r\beta_r+\varphi c_r\alpha_r}{\mu+\varphi}-J_5v_5=0
\end{array}
\right.
\label{systeme_5}
\end{eqnarray}

The left eigenvector is computed and $v_2=0, v_1=0$,and setting $v_3=1$ we get:
\begin{eqnarray}
\left\{
\begin{array}{l}
v_1=0\\
v_2=0\\
v_3=1\\
v_4=\dfrac{\mu+\omega}{\omega}\\
v_5=\dfrac{\mu b_r\beta_r+\varphi c_r\alpha_r}{\mu(\mu+\varphi)}
\end{array}
\right.
\label{systeme_6}
\end{eqnarray}

The theorem of Chavez and Song (2004) is repeated here for convenience, and it may be used to demonstrate the local stability of the endemic equilibrium point around $R_0= 1$.\\

\textbf{Theorem 2}: Consider the following general system of ordinary differential equations with a parameter $\beta^*_i$\\
\begin{eqnarray*}
\dfrac{dx}{dt}=f(x,\beta^*_i),f:\mathbb{R}^n \times \mathbb{R} \rightarrow \mathbb{R} \mbox{  and  } f\in \mathbb{C}^2(\mathbb{R}^n \times \mathbb{R})
\end{eqnarray*}

where $0$ is an equilibrium point of the system, (that is $f (0. \beta^*_i) = 0$ for all $\beta^*_i$ and assume:\\
$A_1:J=D_x f(0,0) = \dfrac{df_i}{dx_i}(0,0)$ is the linearization matrix of System \eqref{matrice5} around the equilibrium point. Zero is a simple eigenvalue of $A$ and all other eigenvalues
of $J$ have negative real parts;\\
$A2$ : Matrix $J$ has a nonnegative right eigenvector $w$ and a left eigenvector $v$ corresponding to the zero eigenvalue.\\

Let $f_k$ be the $k^{th}$ component of $f$ and
\begin{eqnarray}
a=\sum^{n}_{k,i,j=1}v_k w_iw_j \dfrac{\partial^2f_k}{\partial x_i\partial x_j} (0,0)\\
b=\sum^{n}_{k,i=1} v_k w_i \dfrac{\partial^2f_k}{\partial x_i\partial\beta^*_i} (0,0)
\end{eqnarray}

Then the local dynamics of the system around the equilibrium piont $0$ are totally determined by the sign of $a$ and $b$ .
\begin{enumerate}
    \item $a > 0$, $b> 0$. When $\varphi < 0$ with $|\beta_i^*| << 1$, 0 is locally asymptotically stable, and
there exists a positive unstable equilibrium; when $0 < \beta^*_i << 1$, $0$ is unstable and there exists a negative and locally asymptotically stable equilibrium;
\item  $a < 0$, $b < 0$. When $\beta_i^* < 0$ with $|\beta_i^*| << 1$, $0$ is unstable; when $0 < \beta_i^* << 1$, $0$ is locally asymptotically stable, and there exists a positive unstable equilibrium;
\item $a > 0$, $b < 0$. When $\beta_i^* < 0$ with $|\beta_i^*| << 1$, $0$ is unstable, and there exists a locally asymptotically stable negative equilibrium; when $0 < \beta_i^* << 1$, $0$ is stable, and a positive unstable equilibrium appears;
\item $a < 0$, $b > 0$. When $\beta_i^*$ changes from negative to positive, $0$ changes its stability from stable to unstable. Correspondingly a negative unstable equilibrium becomes positive and locally asymptotically stable.\\
Particularly, if $a > 0$ and $b > 0$, then, a backward bifurcation occurs at $\beta_i^* = 0.$
\end{enumerate}

\textbf{Computation of $a$ and $b$}

Since $v_1=v_2=0$ for $k=1,2$, we consider only $k=3,4,5$, the values of $a$ and $b$ are obtained from:

\begin{eqnarray}
a&=& \sum_{i,j=1}^{5}v_3w_iw_j\dfrac{\partial^2f_3}{\partial x_i\partial x_j}+\sum_{i,j=1}^{5}v_4w_iw_j\dfrac{\partial^2f_4}{\partial x_i\partial x_j}+\sum_{i,j=1}^{5}v_5w_iw_j\dfrac{\partial^2f_5}{\partial x_i\partial x_j}\label{a}\\
b&=&\sum_{i=1}^{5}v_3w_i\dfrac{\partial ^2f_3}{\partial x_i \partial \beta_i^*}+\sum_{i=1}^{5}v_4w_i\dfrac{\partial ^2f_4}{\partial x_i \partial \beta_i^*}+\sum_{i=1}^{5}v_5w_i\dfrac{\partial ^2f_5}{\partial x_i \partial \beta_i^*}
\label{b}
\end{eqnarray}

The corresponding non-zero second order partial derivatives at disease-free equilibrium for system \eqref{Systemwithx} are provided by:
\begin{eqnarray}
\dfrac{\partial^2f_3}{\partial x_1\partial x_4}&=&\dfrac{\varphi\mu}{p(\mu+\varphi)}(b_i\beta_i-c_i\alpha_i)\label{f3x1x4}\\
\dfrac{\partial^2f_3}{\partial x_1\partial x_5}&=&\dfrac{\varphi\mu}{p(\mu+\varphi)}(b_r\beta_r-c_r\alpha_r)\label{f3x1x5}\\
\dfrac{\partial^2f_3}{\partial x_2\partial x_4}&=&-\dfrac{\mu^2}{p(\mu+\varphi)}(b_i\beta_i-c_i\alpha_i)\label{f3x2x4}\\
\dfrac{\partial^2f_3}{\partial x_2\partial x_5}&=&-\dfrac{\mu^2}{p(\mu+\varphi)}(b_r\beta_r-c_r\alpha_r)\label{f3x2x5}\\
\dfrac{\partial^2f_3}{\partial x_4\partial x_5}&=&-\dfrac{\mu}{p(\mu+\varphi)}(\mu b_i\beta_i-\varphi c_i\alpha_i)-(\mu b_r\beta_r-\varphi c_r\alpha_r)\label{f3x4x5}\\
\dfrac{\partial^2f_3}{\partial x_4\partial \beta^*_i}&=&\dfrac{\mu}{\mu+\varphi}b_i \label{f3b}
\end{eqnarray}

By replacing \eqref{f3x1x4}, \eqref{f3x1x5}, \eqref{f3x2x4}, \eqref{f3x2x5}, \eqref{f3x4x5} in $a$ \eqref{a}

\begin{eqnarray*}
a &=& 2w_1w_4\dfrac{\partial ^2 f_3}{\partial x_1 \partial x_4}+2w_1w_5\dfrac{\partial ^2 f_3}{\partial x_1 \partial x_5}+2w_2w_4\dfrac{\partial ^2 f_3}{\partial x_2 \partial x_4}+2w_2w_5\dfrac{\partial ^2 f_3}{\partial x_2 \partial x_5}+2w_4w_5\dfrac{\partial ^2 f_3}{\partial x_4 \partial x_5}\\
&=& -\dfrac{2\mu^2\varphi(\gamma b_i\beta_i+\mu b_r\beta_r)}{p(\mu+\varphi)^3}(b_i\beta_i-c_i\alpha_i)-\dfrac{2\varphi\mu(\gamma b_i \beta_i+\mu b_r\beta_r)}{p(\mu+\varphi)^3}(b_r\beta_r-c_r\alpha_r)\\
& &+\dfrac{2\varphi\mu^2 [\mu c_i \alpha_i+\gamma c_r\alpha_r-\gamma(\mu+\varphi)^2]}{\gamma^2p(\mu+\varphi)^2}(b_r\beta_r-c_r\alpha_r)\\
& &-\dfrac{2\mu^2}{\gamma p(\mu+\varphi)}[(\mu b_i\beta_i+\varphi c_i \alpha_i)+\mu(b_r\beta_r+\varphi c_r \alpha_r)]\\
&=&-2(\gamma b_i\beta_i+\mu b_r\beta_r)\dfrac{\varphi\mu}{p(\mu+\varphi)^3}\Big[\dfrac{\mu}{\gamma}(b_i\beta_i-c_i\alpha_i)+(b_r\beta_r-c_r\alpha_r)\Big]\\
& &- 2\mu\varphi\Big[\dfrac{\mu c_i\alpha_i+\gamma c_r\alpha_r-\gamma(\mu+\varphi)^2}{\gamma p(\mu+\varphi)^2}\Big]\Big[\dfrac{\mu}{\gamma}(b_i\beta_i-c_i\alpha_i)+(b_r\beta_r-c_r\alpha_r)\Big]\\
& &-\dfrac{2\mu^2}{\gamma p(\mu+\varphi)}[(\mu b_i\beta_i+\varphi c_i\alpha_i)+(\mu b_r\beta_r+\varphi c_r\alpha_r)]\\
&=&\dfrac{2\varphi\mu^2}{\gamma p(\mu+\varphi)^3}c_i\alpha_i(\gamma b_i\beta_i+\mu b_r\beta_r)+\dfrac{2\varphi\mu}{
 p(\mu+\varphi)^3}c_r\alpha_r(\gamma b_i\beta_i+\mu b_r\beta_r)\\
 & &+\dfrac{2\varphi\mu^2}{\gamma^2
 p(\mu+\varphi)^2}b_i\beta_i(\mu c_i\alpha_i+\gamma c_r\alpha_r)
 +\dfrac{2\varphi\mu}{\gamma
 p(\mu+\varphi)^2}b_r\beta_r(\mu c_i\alpha_i+\gamma c_r\alpha_r)+\dfrac{2\varphi\mu^2}{\gamma
 p} c_i\alpha_i+\dfrac{2\varphi\mu}{ p} c_r\alpha_r \\
 &&-\dfrac{2\varphi\mu^2}{\gamma p(\mu+\varphi)^3}b_i\beta_i(\gamma b_i\beta_i+\mu b_r\beta_r)
 -\dfrac{2\varphi\mu}{
 p(\mu+\varphi)^3}b_r\beta_r(\gamma b_i\beta_i+\mu b_r\beta_r)\\
 && -\dfrac{2\varphi\mu^2}{\gamma^2
 p(\mu+\varphi)^2}c_i\alpha_i(\mu c_i\alpha_i+\gamma c_r\alpha_r)-\dfrac{2\varphi\mu}{\gamma
 p(\mu+\varphi)^2}c_r\alpha_r(\mu c_i\alpha_i+\gamma c_r\alpha_r)-\dfrac{2\varphi\mu^2}{\gamma
 p} b_i\beta_i\\
 &&-\dfrac{2\varphi\mu}{ p} b_r\beta_r-\dfrac{2\mu^2}{\gamma p(\mu+\varphi)}[(\mu b_i\beta_i+\varphi c_i\alpha_i)+(\mu b_r\beta_r+\varphi c_r\alpha_r)]\\
 &=&\dfrac{2\varphi\mu}{\gamma p(\mu+\varphi)^3}(\gamma b_i\beta_i+\mu b_r\beta_r)(\mu c_i\alpha_i+\gamma c_r\alpha_r)+\dfrac{2\varphi\mu}{\gamma^2 p(\mu+\varphi)^2}(\mu c_i\alpha_i+\gamma c_r\alpha_r)(\mu b_i\beta_i+\gamma b_r\beta_r)\\
& &+\dfrac{2\varphi\mu}{\gamma p}(\mu c_i\alpha c_i+\gamma c_r\alpha_r)-\dfrac{2\varphi\mu}{\gamma p(\mu+\varphi)^3}(\gamma b_i\beta_i+\mu b_r\beta_r)(\mu b_i\beta_i+\gamma b_r\beta_r)\\
& &-\dfrac{2\varphi\mu}{\gamma^2 p(\mu+\varphi)^2}(\mu c_i\alpha_i+\gamma c_r\alpha_r)^2-\dfrac{2\varphi\mu}{\gamma p}(\mu b_i\beta_i+\gamma b_r\beta_r)\\
& &-\dfrac{2\mu^2}{\gamma p(\mu+\varphi)}[(\mu b_i\beta_i+\varphi c_i\alpha_i)+(\mu b_r\beta_r+\varphi c_r\alpha_r)]
\end{eqnarray*}


So, $a=A-B$ with  

\begin{eqnarray*}
A&=&\dfrac{2\varphi\mu}{\gamma p(\mu+\varphi)^3}(\gamma b_i\beta_i+\mu b_r\beta_r)(\mu c_i\alpha_i+\gamma c_r\alpha_r)+\dfrac{2\varphi\mu}{\gamma^2 p(\mu+\varphi)^2}(\mu c_i\alpha_i+\gamma c_r\alpha_r)(\mu b_i\beta_i+\gamma b_r\beta_r)\\
& &+\dfrac{2\varphi\mu}{\gamma p}(\mu c_i\alpha c_i+\gamma c_r\alpha_r)\\
B&=&\dfrac{2\varphi\mu}{\gamma p(\mu+\varphi)^3}(\gamma b_i\beta_i+\mu b_r\beta_r)(\mu b_i\beta_i+\gamma b_r\beta_r)-\dfrac{2\varphi\mu}{\gamma^2 p(\mu+\varphi)^2}(\mu c_i\alpha_i+\gamma c_r\alpha_r)^2\\
& &-\dfrac{2\varphi\mu}{\gamma p}(\mu b_i\beta_i+\gamma b_r\beta_r)-\dfrac{2\mu^2}{\gamma p(\mu+\varphi)}[(\mu b_i\beta_i+\varphi c_i\alpha_i)+(\mu b_r\beta_r+\varphi c_r\alpha_r)]
\end{eqnarray*}

\vspace*{1cm}

\hspace{4cm}\textbf{Possibilities of the sign of a}

\vspace*{1cm}

\begin{itemize}
\item Case 1: if $A>B$, then $a>0$
\item Case 2: if $A <B $, then $a<0$
\end{itemize}

By replacing \eqref{f3b} in $b$ \eqref{b}

\begin{eqnarray*}
b&=&2v_3w_4\dfrac{\partial^2 f_3}{\partial x_4\partial \beta_i^*}\\
&=&\dfrac{2\mu^2}{\gamma(\mu+\varphi)}b_i > 0
\end{eqnarray*}

\textbf{Lemma} If $a>0$ then the model~\eqref{model_System1} undergoes a backward bifurcation at $R_0$ close to $1$, if not then the endemic equilibrium point is locally asymptotically stable for $R_0>1$.\\

In addition, when $R_0<1$, the  existence of a backward bifurcation means that there is a potential coexistence of an endemic equilibrium and a disease-free equilibrium. In this case, the strategy of reducing the basic reproduction number to a value less than unity would not be sufficient for the eradication of Ebola disease even with high coverage of vaccination. Such situation will have an impact on herd immunity as we will see below.

\section{Discussion}

In this paper an ordinardinary differential equation (ODE) model for the transmission of EVD, with 5 variables(S,V,E,I,R) was analyzed. The model takes into account a changing total human population over the time, it includes new recruits in the susceptible class. We proved that the model is mathematically and epidemiologically well posed, we also proved the existence of the disease free equilibrium  and endemic equilibrium points. The basic reproduction number, $R_0$, was used to assess the stability of the disease-free steady state and the presence of endemic equilibria.

We showed that when $R_0 < 1$, then the DFE is locally asymptotically stable and unstable when $R_0 > 1$. Our model undergoes a backward bifurcation at $R_0$ close to $ 1 $ which shows a possible cohabitation between the EE and the DFE, In general this shows us that it can not be possible to prove that the EE is unique and stable for $R_0>1$. We noticed that $R_0$ cannot be considered as a modified control intervention measure in our model regardless of $R_0<1$, the disease will not be eradicated.












\bibliographystyle{unsrt}
\bibliography{ebola.bib}







\end{document}